\documentclass[twocolumn,prl,aps,superscriptaddress]{revtex4-1}

\usepackage{graphicx}
\usepackage{amsmath}
\usepackage{amssymb}
\usepackage{marvosym}
\usepackage{txfonts}
\usepackage{bm}
\usepackage{xcolor}
\usepackage{color}
\usepackage{ulem}
\usepackage{makecell}
\usepackage{multirow}
\usepackage[colorlinks=true,urlcolor=blue,anchorcolor=blue,linkcolor=blue,citecolor=blue,breaklinks=true]{hyperref}
 
\newcommand{\aov}{$\mathrm{V}_\mathrm{O}$}

\begin{document}

\normalem




\textbf{Comment on ``Anisotropic Scattering Caused by Apical Oxygen Vacancies in Thin Films of Overdoped High-Temperature Cuprate Superconductors"} In Ref.~\onlinecite{Wang2022}, Wang \textit{et al}.\ address an important problem in the overdoped cuprates, taking the first steps toward understanding the structure of real defects in these materials, in this case the apical oxygen vacancy, \aov, that sits immediately above the planar Cu.  However, the approach taken has some issues, which we outline in this Comment.

\begin{figure}[t]
    \centering
    \includegraphics[width = 0.85 \linewidth]{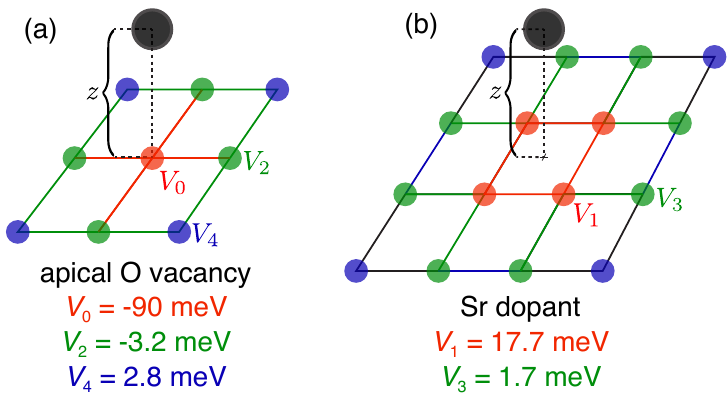}
    \caption{Defects in La$_{2-x}$Sr$_x$CuO$_4$ showing \textit{ab-initio} impurity potentials at surrounding Cu sites, from Ref.~\onlinecite{Ozdemir:2022}. (a) The site-centered apical oxygen vacancy is nearly pointlike, with $V_0 \gg V_2, V_4$. (b) The plaquette-centered Sr dopant is inherently extended in real space.}
    \label{fig:impuritiesLSCO}
\end{figure}

\textit{Impurity potentials:} The primary assumption of Ref.~\onlinecite{Wang2022}, based on symmetry arguments, is that since
there is no hopping between the apical oxygen $p_z$ orbital and the Cu~$3d_{x^2 - y^2}$ orbital immediately below it, the leading couplings from the apical oxygen are to the next-nearest-neighboring Cu sites.  This leads them to a real-space scattering potential for the vacancy that has no on-site term, and hence scattering matrix elements $V_{\mathbf{k},\mathbf{k}'}$ that depend strongly on momentum. However, when \textit{ab-initio} calculations are carried out, as in Ref.~\onlinecite{Ozdemir:2022}, the impurity potential of the site-centered apical oxygen vacancy in LSCO is in fact \textit{dominated} by the on-site term, to the point that the \aov\ defect essentially acts as a point scatterer (see Fig.~\ref{fig:impuritiesLSCO}(a)), with matrix elements that are almost independent of momentum transfer $\mathbf{q}$.  The reason for this discrepancy is that by making symmetry arguments analogous to those for hopping processes, Ref.~\onlinecite{Wang2022} overlooks the strong electrostatic contribution the \aov\ defect makes to the energy of the Cu site immediately below it. Nevertheless, there does exist a source of extended impurity potentials in LSCO --- the plaquette-centered Sr dopants (Fig.~\ref{fig:impuritiesLSCO}(b)) ---  which give rise to strongly momentum-dependent $V_{\mathbf{k},\mathbf{k}'}$. 

\textit{Fermi surface:} Overdoped LSCO undergoes a Lifshitz transition at which the van Hove singularity passes through the Fermi level \cite{Yoshida:2006hw}.  As a result, the  Fermi surface is far from isotropic, and the density of states and Fermi velocity vary strongly with angle and doping, in contrast to the circular Fermi surface assumed in Ref.~\onlinecite{Wang2022}.  This anisotropy has important consequences for physical properties such as superfluid density \cite{Lee-Hone:2017} and conductivity~\cite{Lee-Hone:2018}.  Fermi-surface anisotropy must also be present for impurity scattering to produce a strongly angle-dependent scattering rate $\Gamma_\theta$, a key assumption in Ref.~\onlinecite{Wang2022}.

\textit{Self energies:} The rich complexity of dirty \mbox{$d$-wave} superconductors stems from the nontrivial and often surprising ways in which disorder alters almost all physical properties.  The effects of disorder enter via the self energy, which must be calculated carefully. In Ref.~\onlinecite{Wang2022}, a heuristic scattering rate $\Gamma_\theta$ is used to renormalize the quasiparticle energies via $\omega \to \omega + i \Gamma_\theta$.  This approach can work in the normal state, but breaks down in a $d$-wave superconductor, where disorder leads to self energies with nontrivial energy dependence.  Physically speaking, the disorder in a $d$-wave superconductor breaks pairs and thereby modifies the energy-dependent density of states, which alters the phase space for recoil, which in turn changes the scattering rate and therefore the amount of pair-breaking.  The whole process must be treated self-consistently, as a function of energy.  In addition, for the types of momentum-dependent scattering (i.e., non-pointlike scattering) of interest in LSCO, not only must the quasiparticle energy be renormalized, but also the \mbox{$d$-wave} gap.  In Ref.~\onlinecite{Wang2022} there is no self-consistent treatment of the self energies and no explicit gap renormalization: the substitution $\omega \to \omega + i \Gamma_\theta$ is simply inserted into the clean-limit expressions for gap equation, superfluid density and conductivity.

\textit{Vertex corrections:} Finally, the types of momentum-dependent potentials considered in Ref.~\onlinecite{Wang2022} have strong forward-scattering character, which enhances the electrical current.  As a result, vertex corrections (small-angle scattering corrections) must be included in the calculation of two-particle properties such as superfluid density and optical conductivity, but this has been neglected.

For a treatment of the disorder problem in overdoped cuprates that explains these subtleties, see Ref.~\onlinecite{Ozdemir:2022}.\\

\noindent H.~U.~\"Ozdemir,$^1$ Vivek Mishra,$^2$ N.~R.~Lee-Hone,$^1$ Xiangru Kong,$^3$ T.~Berlijn,$^3$ D.~M.~Broun,$^1$ and P.~J.~Hirschfeld$^4$\\
{\small
\indent $^1$Department of Physics, Simon Fraser University, Burnaby, BC, V5A 1S6, Canada\\
\indent $^2$Kavli Institute for Theoretical Sciences, University of Chinese Academy of Sciences, Beijing 100190, China\\
\indent $^3$Center For Nanophase Materials Sciences, Oak Ridge National Laboratory, Oak Ridge, TN 37831, USA\\
\indent $^4$Department of Physics, University of Florida, Gainesville FL 32611
}


\begin{thebibliography}{5}%
\makeatletter
\providecommand \@ifxundefined [1]{%
 \@ifx{#1\undefined}
}%
\providecommand \@ifnum [1]{%
 \ifnum #1\expandafter \@firstoftwo
 \else \expandafter \@secondoftwo
 \fi
}%
\providecommand \@ifx [1]{%
 \ifx #1\expandafter \@firstoftwo
 \else \expandafter \@secondoftwo
 \fi
}%
\providecommand \natexlab [1]{#1}%
\providecommand \enquote  [1]{``#1''}%
\providecommand \bibnamefont  [1]{#1}%
\providecommand \bibfnamefont [1]{#1}%
\providecommand \citenamefont [1]{#1}%
\providecommand \href@noop [0]{\@secondoftwo}%
\providecommand \href [0]{\begingroup \@sanitize@url \@href}%
\providecommand \@href[1]{\@@startlink{#1}\@@href}%
\providecommand \@@href[1]{\endgroup#1\@@endlink}%
\providecommand \@sanitize@url [0]{\catcode `\\12\catcode `\$12\catcode
  `\&12\catcode `\#12\catcode `\^12\catcode `\_12\catcode `\%12\relax}%
\providecommand \@@startlink[1]{}%
\providecommand \@@endlink[0]{}%
\providecommand \url  [0]{\begingroup\@sanitize@url \@url }%
\providecommand \@url [1]{\endgroup\@href {#1}{\urlprefix }}%
\providecommand \urlprefix  [0]{URL }%
\providecommand \Eprint [0]{\href }%
\providecommand \doibase [0]{http://dx.doi.org/}%
\providecommand \selectlanguage [0]{\@gobble}%
\providecommand \bibinfo  [0]{\@secondoftwo}%
\providecommand \bibfield  [0]{\@secondoftwo}%
\providecommand \translation [1]{[#1]}%
\providecommand \BibitemOpen [0]{}%
\providecommand \bibitemStop [0]{}%
\providecommand \bibitemNoStop [0]{.\EOS\space}%
\providecommand \EOS [0]{\spacefactor3000\relax}%
\providecommand \BibitemShut  [1]{\csname bibitem#1\endcsname}%
\let\auto@bib@innerbib\@empty
\bibitem [{\citenamefont {Wang}\ \emph {et~al.}(2022)\citenamefont {Wang},
  \citenamefont {Xu}, \citenamefont {Zhang},\ and\ \citenamefont
  {Wang}}]{Wang2022}%
  \BibitemOpen
  \bibfield  {author} {\bibinfo {author} {\bibfnamefont {D.}~\bibnamefont
  {Wang}}, \bibinfo {author} {\bibfnamefont {J.-Q.}\ \bibnamefont {Xu}},
  \bibinfo {author} {\bibfnamefont {H.-J.}\ \bibnamefont {Zhang}}, \ and\
  \bibinfo {author} {\bibfnamefont {Q.-H.}\ \bibnamefont {Wang}},\ }\href
  {\doibase 10.1103/PhysRevLett.128.137001} {\bibfield  {journal} {\bibinfo
  {journal} {Phys. Rev. Lett.}\ }\textbf {\bibinfo {volume} {128}},\ \bibinfo
  {pages} {137001} (\bibinfo {year} {2022})}\BibitemShut {NoStop}%
\bibitem [{\citenamefont {{\"O}zdemir}\ \emph {et~al.}(2022)\citenamefont
  {{\"O}zdemir}, \citenamefont {Mishra}, \citenamefont {Lee-Hone},
  \citenamefont {Kong}, \citenamefont {Berlijn}, \citenamefont {Broun},\ and\
  \citenamefont {Hirschfeld}}]{Ozdemir:2022}%
  \BibitemOpen
  \bibfield  {author} {\bibinfo {author} {\bibfnamefont {H.~U.}\ \bibnamefont
  {{\"O}zdemir}}, \bibinfo {author} {\bibfnamefont {V.}~\bibnamefont {Mishra}},
  \bibinfo {author} {\bibfnamefont {N.~R.}\ \bibnamefont {Lee-Hone}}, \bibinfo
  {author} {\bibfnamefont {X.}~\bibnamefont {Kong}}, \bibinfo {author}
  {\bibfnamefont {T.}~\bibnamefont {Berlijn}}, \bibinfo {author} {\bibfnamefont
  {D.~M.}\ \bibnamefont {Broun}}, \ and\ \bibinfo {author} {\bibfnamefont
  {P.~J.}\ \bibnamefont {Hirschfeld}},\ }\href@noop {} {} (\bibinfo {year}
  {2022}),\ \Eprint {http://arxiv.org/abs/2206.01348} {arXiv:2206.01348}
  \BibitemShut {NoStop}%
\bibitem [{\citenamefont {Yoshida}\ \emph {et~al.}(2006)\citenamefont {Yoshida}
  \emph {et~al.}}]{Yoshida:2006hw}%
  \BibitemOpen
  \bibfield  {author} {\bibinfo {author} {\bibfnamefont {T.}~\bibnamefont
  {Yoshida}} \emph {et~al.},\ }\href {\doibase 10.1103/PhysRevB.74.224510}
  {\bibfield  {journal} {\bibinfo  {journal} {Phys.\ Rev.\ B}\ }\textbf
  {\bibinfo {volume} {74}},\ \bibinfo {pages} {224510} (\bibinfo {year}
  {2006})}\BibitemShut {NoStop}%
\bibitem [{\citenamefont {Lee-Hone}\ \emph {et~al.}(2017)\citenamefont
  {Lee-Hone}, \citenamefont {Dodge},\ and\ \citenamefont
  {Broun}}]{Lee-Hone:2017}%
  \BibitemOpen
  \bibfield  {author} {\bibinfo {author} {\bibfnamefont {N.~R.}\ \bibnamefont
  {Lee-Hone}}, \bibinfo {author} {\bibfnamefont {J.~S.}\ \bibnamefont {Dodge}},
  \ and\ \bibinfo {author} {\bibfnamefont {D.~M.}\ \bibnamefont {Broun}},\
  }\href {\doibase 10.1103/PhysRevB.96.024501} {\bibfield  {journal} {\bibinfo
  {journal} {Phys. Rev. B}\ }\textbf {\bibinfo {volume} {96}},\ \bibinfo
  {pages} {024501} (\bibinfo {year} {2017})}\BibitemShut {NoStop}%
\bibitem [{\citenamefont {Lee-Hone}\ \emph {et~al.}(2018)\citenamefont
  {Lee-Hone}, \citenamefont {Mishra}, \citenamefont {Broun},\ and\
  \citenamefont {Hirschfeld}}]{Lee-Hone:2018}%
  \BibitemOpen
  \bibfield  {author} {\bibinfo {author} {\bibfnamefont {N.~R.}\ \bibnamefont
  {Lee-Hone}}, \bibinfo {author} {\bibfnamefont {V.}~\bibnamefont {Mishra}},
  \bibinfo {author} {\bibfnamefont {D.~M.}\ \bibnamefont {Broun}}, \ and\
  \bibinfo {author} {\bibfnamefont {P.~J.}\ \bibnamefont {Hirschfeld}},\ }\href
  {\doibase 10.1103/PhysRevB.98.054506} {\bibfield  {journal} {\bibinfo
  {journal} {Phys. Rev. B}\ }\textbf {\bibinfo {volume} {98}},\ \bibinfo
  {pages} {054506} (\bibinfo {year} {2018})}\BibitemShut {NoStop}%
\end{thebibliography}

%

\clearpage

\end{document}